\documentclass[preprint,showpacs,amsmath]{revtex4}
\usepackage[dvips]{graphicx}
\usepackage{comment}
\usepackage{amsmath}
\usepackage{xspace}
\usepackage{ulem}
\usepackage{subfigure}
\usepackage{placeins}
\usepackage{cancel}
\usepackage{color}

\newcommand{\elabel}[1]{\label{eq:#1}}

\begin{document}

\title{
Testing Universality in Critical Exponents:
the Case of Rainfall
}

\author{Anna Deluca$^{1}$, Pedro Puig$^2$, \'Alvaro Corral$^{3,2}$}

\affiliation{
$^1$ Max Planck Institute for the Physics of Complex Systems, Noethnitzer Str. 38, D-01187 Dresden, Germany \\
$^2$ Departament de Matem\`atiques, Universitat Aut\`onoma de Barcelona, E-08193 Cerdanyola del Vall\'es, Spain\\
$^3$ Centre de Recerca Matem\`atica, Campus de Bellaterra, Edifici C, E-08193 Barcelona, Spain\\
}
\date{\today}
\begin{abstract}
One of the key clues to consider rainfall as a self-organized critical 
phenomenon is the existence of power-law distributions for 
rain-event sizes.
We have studied the problem of universality in the exponents of
these distributions by means of a suitable statistic whose
distribution is inferred by several variations of a permutational test.
In contrast to more common approaches, our procedure 
does not suffer from the difficulties of multiple testing 
and does not require the precise knowledge of the uncertainties associated
to the power-law exponents.
When applied to seven sites monitored by the Atmospheric Radiation Measurement 
Program the tests lead to the rejection of the universality hypothesis, 
despite the fact that the exponents are rather close to each other.
{We discuss the reasons of the rejection.}
\end{abstract}


\maketitle

\section{Introduction}

The concept of universality is
``one of the most striking features of the theory of
critical phenomena'' \cite{Yeomans1992}, 
giving sense to the extended use of modeling
in statistical physics.
Strictly, it would mean that using a naive model
one could derive all the critical exponents
and scaling functions of any real system
displaying a second order phase transition,
no matter the complications of the 
interactions in the system, 
because critical exponents and 
scaling functions should be ``universal''.
In practice, the name universality is 
somewhat pretentious \cite{Stanley_rmp}, and what one instead 
obtains are several ``universality classes'',
which are sets of systems sharing
the same behavior (in terms of
critical exponents and scaling functions), 
depending only, in equilibrium and for systems with short-range interactions, 
on the dimensionality of space
and the symmetry of the order parameter.
This is in sharp contrast with the behavior of other
important properties, as for instance the critical 
temperature, justifying the perplexity for the 
universality phenomenon.

The classification of many disparate systems 
into a relatively reduced number of universality classes
is therefore a fundamental problem (analogous somehow to the 
construction of the Mendeleev's periodic table \cite{Stanley_rmp}), 
which relies on the accurate determination of critical exponents
and scaling functions.
A weaker form of universality considers only 
the coincidence of the critical exponents, disregarding
the scaling functions.
This is due, when dealing with experimental or numerical 
data, to the fact that critical exponents
can be obtained directly as a single number each one,
whereas scaling functions need to be parameterized
(which introduces some degree of arbitrariness in the
parameterization);
otherwise, scaling functions need to be obtained from the analytical solution of a model.


Among all the different critical exponents, an important subset
are those arising from probability distributions, 
such as cluster number densities, avalanche size distributions, etc.
{\cite{Aharony,Christensen_Moloney}}.
In this case, the probability mass function or the probability density 
$f(s)$ of the variable $s$ can be written, 
at least for large $s$, as
\begin{equation}
f(s) = s^{-\tau} G(s/s_c),
\label{scaling_eq}
\end{equation}
%
where $s_c$ is a characteristic value of $s$
and $G$ is a scaling function that can be an exponential or any other 
function going to a constant for small $s$
and decaying very fast for large $s$.
Close to the critical point
{and in the infinite system-size limit}, $s_c$ diverges,
$G$ tends to a constant, $f(s)$ becomes a power law, and $\tau$ emerges as 
a genuine critical exponent.

Similar situations arise outside critical phenomena;
for instance, in anomalous diffusion,
the long-term behavior of a diffusion process 
(with short-range correlations) can be classified
within a continuous of universality classes 
defined by the L\'evy-stable laws,
characterized by power-law tails \cite{Bouchaud_Georges}.
Although the behavior of the system is not governed
by a continuous phase transition, it is possible
to understand it from the existence of a fixed point
in some renormalization-group transformation equations \cite{Corral_csf}.
Other stochastic processes lead to analogous situations 
\cite{Corral_csf,Gyorgyi,Corral_jstat}.

The determination of critical exponents is not an easy task, 
even more difficult when they are the exponents of power-law distributions.
Very recently, considerable attention has been devoted 
to the proper fit of such distributions,
together with the subsequent goodness-of-fit testing.
{White et al.} and
Clauset et al. (among others) \cite{White,ClausetShaliziNewman2009} mention the systematic errors that
can arise from using the least-square linear regression method
applied to $\ln f(s)$ {as a function of $\ln s$},
although the alternative recipe proposed by Clauset et al.
to find the {most suitable} power-law range has been found 
to perform badly in some cases \cite{CorralETAL2011},
so somewhat different methods have been {suggested} by other authors
\cite{PetersDelucaETAL2010,Corral_Deluca}.

But determining the critical exponents as accurately and unbiasedly 
as possible, together with their associated uncertainties, 
is not the end of the story if one is looking for universality.
The exponents need to be properly compared, in order to test if they
are statistically compatible {with each other} {or not}.
{From a more practical point of view,
if universality does not hold,  
one may monitor 
some process by the changes in the value
of some power-law exponent, 
which can play the role of a precursor
of catastrophic failure
(see citations at Ref. \cite{Baro_Corral}).
}

{{The subject of this paper is to develop a systematic procedure to 
compare critical exponents, applying it to}
study in detail a non-equilibrium problem: that of universality in 
rain-event size distributions,
which is important to characterize rainfall as a self-organized critical phenomena
\cite{Christensen_Moloney,Bak_book}.
These distributions were first analyzed for one single location in the Baltic coast
by Peters et al. \cite{Peters_prl,Peters_pre},
who reported a power-law distribution with an exponent $\tau$ around 1.4.
More recently, Ref. \cite{PetersDelucaETAL2010} widened the study 
to 10 sites around the globe; 
after discarding 3 of them due to different
instrumentally induced biases and errors, 
not only the power-law hypothesis was confirmed 
but also the scaling form of the distribution, Eq. (\ref{scaling_eq}),
with {rather smaller} exponents, ranging from $\tau=1.14$ to 1.19.
However, a proper statistical test to decide if the exponents
were compatible with a unique value or not was not attempted.

This is what we undertake here, extending the study in
order to include new data.
In the following section we introduce the rain data, 
the definition of rain events,
and the precise way of fitting the power-law exponents.
Next, in Sec. \ref{sec_three}, we explain some naive ways to compare the values of the exponents;
{these ways are not satisfactory due to the difficulties of multiple testing, 
and, more important, because they require the precise estimation 
of the uncertainties of the exponents, which may be impossible in practice.}
Section IV is devoted to the development of a simple {and intuitive} permutational test
in order to investigate the universality of the exponents
for the rain data of Ref. \cite{PetersDelucaETAL2010},
whereas in Sec. \ref{sec_five} this test is generalized.
Section VI presents the more complete, improved test, 
as an extension of the previous one. 
We finish with some {discussion and} conclusions.
}


\section{Data, Rain Events, and Power-law fitting}

As in Ref. \cite{PetersDelucaETAL2010},
we analyze rain data from the Atmospheric Radiation Measurement (ARM) Program
(www.arm.gov). 
The {ARM rain} database presents the advantage of its homogeneity 
in the sense that all sites are equipped with the same
type of pluviometer, 
an optical rain
gauge from MiniOrg (Optical Scientific, Inc.), model ORG-815-DA 
\cite{PetersDelucaETAL2010}.
Rain rate is recorded with a one-minute temporal resolution,
with a minimum value of 0.001 mm/hour,
but we disregard rain rates below 0.2 mm/hour
(i.e., we treat them as zero), 
as recommended by the ARM Handbook.
Other corrections were applied to the data
using the ARM Data Quality Reports. The data measurements have been also carefully compared 
with  measurements from other devices in order to detect and reduce systematic biases \cite{Poster}. 

{In order to compare with the results of Ref. \cite{PetersDelucaETAL2010},}
we consider 
the same sites studied there,
{except} the 3 sites that those authors found problematic 
for diverse reasons
(North Slope of Alaska, Point Reyes, and Southern Great Plains, all 3 in USA); 
this yields a remainder of M=7 sites,
see Table \ref{table_data}.
For some of these sites (Manus, Nauru, Darwin, and Graciosa)
new data are available since the study of Ref. \cite{PetersDelucaETAL2010}, 
so our database has been updated accordingly.
The rest of the sites (Niamey, Heselbach, and Shouxian)
remain essentially the same, except perhaps for little operational errors
reported since then {and corrected in our treatment}.  


\begin{widetext}

\begin{table}[ht]
\caption{ARM observation sites used in our analysis of rainfall,
with corresponding starting and ending times and location.
I. stands for island.}
\begin{tabular}{ l rrrr }

\footnotesize{Site} & \footnotesize{start time} & \footnotesize{end time} &   \footnotesize{latitude}&   \footnotesize{longitude} \\
\hline

\footnotesize{Manus I., Papua New Guinea}& \footnotesize{2005/02/15} & \footnotesize{2012/03/18}  & \footnotesize{ 2.116$^\circ$S} & \footnotesize{147.425$^\circ$E}\\  
\footnotesize{Nauru I.,  Nauru Republic} & \footnotesize{2005/02/15} & \footnotesize{2012/03/18}  & \footnotesize{ 0.521$^\circ$S} & \footnotesize{166.916$^\circ$E}\\
\footnotesize{Darwin, Australia}	   & \footnotesize{2005/02/15} & \footnotesize{2012/03/18}  & \footnotesize{12.425$^\circ$S} & \footnotesize{130.892$^\circ$E}\\ 
 \footnotesize{Niamey, Niger}		   & \footnotesize{2005/12/26} & \footnotesize{2006/12/08}  & \footnotesize{13.522$^\circ$N} & \footnotesize{  2.632$^\circ$E}\\ 
 \footnotesize{Heselbach, Germany}	   & \footnotesize{2007/04/01} & \footnotesize{2008/01/01}  & \footnotesize{48.450$^\circ$N} & \footnotesize{  8.397$^\circ$E}\\ 
\footnotesize{Shouxian, China}	         & \footnotesize{2008/05/09} & \footnotesize{2008/12/28}  & \footnotesize{32.558$^\circ$N} & \footnotesize{116.482$^\circ$E}\\
\footnotesize{Graciosa I., Azores, Portugal}
      					         & \footnotesize{2009/04/14} & \footnotesize{2011/01/06}  & \footnotesize{39.091$^\circ$N} & \footnotesize{ 28.029$^\circ$E}\\
\hline
 \end{tabular}
 \label{table_data} 
\end{table} 
\end{widetext}

{

The fundamental concept in the self-organized-criticality approach is the
rain event, which is defined as a sequence of rain-rate 
values all above a certain threshold 
(starting and ending {just when the threshold is crossed})
\cite{Andrade,Peters_prl,Peters_pre,Deluca_Corral_npg}; in our study the threshold is set to
0.2 mm/hour \cite{PetersDelucaETAL2010},
{although higher thresholds can be also of interest
\cite{Deluca_Moloney_Corral}.}
The size $s$ of the event is the total amount of rain collected
during the lifetime of the event, i.e., 
the time integral of the rain rate {along event duration}.
Rain events containing errors of measurement
are discarded (in contrast to Ref. \cite{PetersDelucaETAL2010},
where the part of the event devoid of errors was counted as an event). 
{A large record of rainfall contains enough events to estimate
the probability density of the rain-event size, $f(s)$,
and, independently, to test if this distribution follows a power law
or not.}

The key to fit properly power-law distributions {to real-world data} is to have
an objective criterion to decide at which point the power law
starts {and (in the truncated case) at which point it ends;} 
these cut-offs define the fitting range. 
{This is so because incompleteness of the data for very small sizes 
and finite-size effects for large sizes lead to considerable deviations
from a power-law regime, see Eq. (\ref{scaling_eq}). As a fitting method}
we essentially use the improvement and extension
of the Clauset et al.'s method \cite{ClausetShaliziNewman2009}
introduced in Ref. \cite{PetersDelucaETAL2010}
and explained in much detail elsewhere \cite{Corral_Deluca}.
Summarizing, 
``all'' fitting ranges are considered, and among those
which yield acceptable fits ({high enough $p-$values}), the one containing more data
{points (i.e., more events)} is selected.

Fitting is performed by maximum likelihood estimation, 
goodness of fit is tested by the Kolmogorov-Smirnov distance,
and the $p-$value of the fit is computed from Monte Carlo simulations.
To be precise, to look for the fitting range we sweep 20 values per order of magnitude
(equidistant in log-scale) of the small-size and large-size cut-offs.
{Then, for each fitting range, the value of the exponent 
is estimated by maximum likelihood,
and the distance between the fit and the empirical data is
quantified by the Kolmogorov-Smirnov statistic.
Monte Carlo simulations are used to generate 300 synthetic 
samples within the fitting range and power-law distributed with the estimated exponent.
The application of the same maximum-likelihood estimation
and Kolmogorov-Smirnov method to each synthetic sample
leads to the distribution of the Kolmogorov-Smirnov distance,
from which the $p-$value of the fit arises.
This is different from}
the uncertainty in the exponent, which can be quantified by the standard
deviation of the maximum likelihood estimation, 
calculated {using the jackknife procedure}
(the formula of Ref. \cite{Aban}, see Eq. (\ref{sigma0}), 
and our Monte Carlo simulations 
are in agreement with this method).
We consider a fit as acceptable (or non-rejectable) if $p > 0.10$.
Among all the acceptable fits, we select the one containing more data points, 
as mentioned above.
%
%
%

Provided that we find at least one non-rejectable fit, for each dataset $i$ 
we end up with three optimized values:
one is the resulting estimated exponent $\bar \tau_i$ and the other two, 
$a_i$ and $b_i$ (the selected small-size and large-size cut-offs), 
{define the fitting} range $a_i \le s \le b_i$ for which
the power-law fit holds.
Notice also that the fitting and testing procedure does
not make use of the estimation of $f(s)$ shown for illustration purposes in the figures of Ref. \cite{PetersDelucaETAL2010}. 
Results of the fits for the rain-event size distributions are shown in Table \ref{table_exponents}.

{{As a first trial,} in order to simplify the comparison between the different sites,
we decide to consider the common range over which all distributions
are power laws. 
We define then $a=\max_{\forall i} a_i$ and $b=\min_{\forall i} b_i$
({verifying that $a \ll b$});
then, new exponents $\hat \tau_i$ are recalculated for this common range just by maximum likelihood
estimation. 
The estimated power-law fittings will be given then by
\begin{equation}
f_i(s) \propto \frac 1 {s^{\hat \tau_i}}, \mbox{ for } a \le s \le b. 
\end{equation}
The resulting exponents will be different but very close to the previous ones
(within the expected fluctuations), see Table \ref{table_exponents}.}
Nevertheless, the $p-$value for the new fits may change, 
even being possible that some of them drop below the acceptance threshold.
This is what naturally happens in goodness-of-fit testing
(as the null hypothesis may be rejected even when it is true).
We do not need to do anything in this regard,
just the reader must be aware of it.


\begin{table}
\caption[]{
Results of power-law fits for the 7 sites studied in Ref. \cite{PetersDelucaETAL2010}
(with updated data).
The total number of rain events (for $0 < s < \infty$) is $N_i$.
The resulting optimum cut-offs $a_i$ and $b_i$ are displayed, in mm, 
together with the resulting number of events in fitting range $\bar n_i$ and exponent $\bar \tau_i$.
When the fits are restricted to the common range, $a=0.0071$ mm and $b=0.501$ mm,
the new number of events and power-law exponents are $n_i$ and $\hat\tau_i$.
The quantity in parenthesis is the standard deviation of the exponents
in terms of the last significant digit of the exponent
{(calculated for the case when the fitting range is fixed a priori)}.
\label{table_exponents}}
\begin{tabular}{l r r r r r r r r}

{Site} $i$ & $N_i$ & $a_i$ & $b_i$ & ${b_i}/{a_i}$ & $\bar n_i$ &   $\bar \tau_i$ & $n_i$ & $\hat \tau_i$ \\
\hline
 \footnotesize{1. Manus} & 15725 &  0.0071 & 10.0 &   1413 & 11910 & 1.152(05) &  8455 & 1.151(09)\\
 \footnotesize{2. Nauru} &  8404 &  0.0063 &  3.2 &    501 &  6350 & 1.120(07) &  4831 & 1.122(12)\\
\footnotesize{3. Darwin} &  5216 &  0.0063 &  3.5 &    562 &  3946 & 1.106(09) &  2959 & 1.095(15)\\
 \footnotesize{4. Niamey} &  260 &  0.0040 & 56.2 &  14125 &   231 & 1.193(26) &   135 & 1.231(72)\\
\footnotesize{5. Heselbach}&2437 &  0.0040 &  0.6 &    141 &  1844 & 1.132(16) &  1569 & 1.149(21)\\
\footnotesize{6. Shouxian} & 476 &  0.0040 &  1.3 &    316 &   372 & 1.165(32) &   290 & 1.185(48)\\
\footnotesize{7. Graciosa}& 4260 &  0.0071 &  0.5 &     71 &  2841 & 1.147(15) &  2841 & 1.147(15)\\
\hline

 \end{tabular}
\end{table}

}

\section{
Difficulties of testing
}
\label{sec_three}


{This section discusses why traditional approaches present difficulties 
when applied to our case, and justifies why other approaches will 
be more convenient. Our choice will be the use of permutational tests, 
introduced in the following sections.
As a starting point,
let us} consider 
the simple case in which one only has to decide if
some exponent (or in general, some statistic) 
$\tau$ takes the same value
or not in two different systems, 1 and 2.
The null hypothesis is then $\tau_1=\tau_2$.
What one usually has is an estimation for each exponent, denoted
as $\hat \tau_i$, with $i=1,2$,
together with an estimation of their standard deviations, 
which, if the number of data for each system is large, we can assume
converges to the true standard deviation, $\sigma_i$.

Under the null hypothesis, the difference $d=\hat \tau_1-\hat \tau_2$
will have zero mean, and, if {datasets 1 and 2 are independent samples}
(which will be the common situation {if the two systems are unrelated}), 
the standard deviation of
the difference of the estimators will be $\sigma_d=\sqrt{\sigma_1^2+\sigma_2^2}$.
As, asymptotically, 
$\hat \tau_1$ and $\hat \tau_2$ are normally distributed \cite{Aban}, 
so will be their difference, 
and therefore it is straightforward to obtain a confidence interval 
for it. If the interval, centered at zero, includes the observed value
of the difference, 
the null hypothesis cannot be rejected 
and the exponents can be considered to take the same value in both systems
(which can belong then to the same universality class, 
at least regarding the exponent $\tau$).
{For instance, if we compare the exponents of the Manus and Darwin sites (sites 1 and 3), the difference between them is
$d=\lvert  {\hat\tau_{Manus}-\hat\tau_{Darwin}}  \rvert=0.055$ with standard deviation  
$\sigma_{d}=\sqrt{\sigma_{Manus}^2+\sigma_{Darwin}^2}=0.0175$, see Table \ref{table_exponents}.
Considering the $1.96 \sigma_d$ interval, we should reject the hypothesis that both exponents are the same, 
with a 95 \% confidence.}
Observe that, although the test for the differences is well known, it is not in agreement
with the {somewhat extended} practice of verifying if the confidence intervals 
of $\hat \tau_1$ and $\hat \tau_2$ overlap,
which yields a smaller significance level
(and is, then, less ``rigorous'', or more permissive
\cite{footnote}).


{But} the situation is not so simple when one needs to analyze 3 or more systems.
Taking the naive approach of comparing the overlap of the confidence intervals,
some systems may lay outside the overlap region of the rest just by chance, 
which will be more likely as the number of systems increases.
{So, the rejection in the previous example could be caused by an unavoidable ``bad luck,''
as those sites are just a part of a much larger collection of sites.}
Also, it might be difficult to define which is the overlap region, 
as {there} can be several subsets, 
or a continuous of overlapping subsets.

{If we take}
all pairs of systems, this leads to
$M(M-1)/2$ {pair} tests if there are $M$ different systems.
{In our case, $M=7$ and $M(M-1)/2=21$.
Comparing the $M$ exponents by pairs
one can see from Table \ref{table_comparison}
that the null hypothesis 
would be rejected in 3 out of the 21 cases, 
at the $5 \%$ significance level.
Are these rejections really significant?
In order to avoid the rejection of the null hypothesis 
for a given test ``by accident'', one can apply  
some correction of the significance level, as the Bonferroni
correction or the \v{S}id\'ak correction
\cite{Bland_Altman,Abdi_Bonferroni}.
With a confidence level of 95 \%
we have a probability of rejecting the null hypothesis when it is true
of {$\alpha\equiv 1-0.95=0.05$
(this is indeed the significance level);} 
so, sooner or later we will get large enough differences
in the exponents, due to statistical fluctuations,
if the number of tests is large enough.
Therefore,
the probability of at least one rejection in 21 independent tests is
$1-(1-\alpha)^{21}=0.67$, and we can consider this number as
the global (or familywise) significance level,
which turns out to be rather high.

\begin{table}
\caption[]{{Absolute} difference between power-law exponents for 
the common fitting range. Uncertainty is evaluated as
$1.96 \sigma_d$, corresponding to $5 \%$ significance level. 
Significant differences are underlined.
{Note that $\sigma_d$ is calculated from the standard deviations 
of the exponents for a fixed fitting {range}.}
\label{table_comparison}
}
\begin{tabular}{l c c c c c c }

 &  \footnotesize{Nauru} &  \footnotesize{Darwin} &  \footnotesize{Niamey} & \footnotesize{Heselbach} &\footnotesize{Shouxian} & \footnotesize{Graciosa} \\
\hline
\footnotesize{Manus}     &  0.03$\pm$0.03 &\underline{0.06$\pm$0.03} &0.08$\pm$0.14 &0.00$\pm$0.04 &0.03$\pm$0.10 &0.00$\pm$0.03\\
\footnotesize{Nauru}     &- &0.03$\pm$0.04 &0.11$\pm$0.14 &0.03$\pm$0.05 &0.06$\pm$0.10 &0.02$\pm$0.04\\
\footnotesize{Darwin}    &- &-&0.14$\pm$0.14 & \underline{0.05$\pm$0.05} &0.09$\pm$0.10 &\underline{0.05$\pm$0.04}\\
\footnotesize{Niamey}    &- &- &- &0.08$\pm$0.15 &0.05$\pm$0.17 &0.08$\pm$0.14\\
\footnotesize{Heselbach} &- &-& -&-&0.04$\pm$0.10 &0.00$\pm$0.05\\
\footnotesize{Shouxian}  &- & -& -&- &- &0.04$\pm$0.10\\
\hline
 \end{tabular}
\end{table}

The idea of the {\v{S}id\'ak} correction is to select $\alpha$
in such a way that the resulting global significance level 
is more reasonable, say 0.05; then, in our case, 
$\alpha=1-\sqrt[21]{1-0.05}=0.0024$.
For the Bonferroni correction one approximates
$1-(1-\alpha)^{21} \simeq 1-(1-21 \alpha)=0.05$,
which leads in this case to essentially the same $\alpha=0.0024$.
This has the advantage of not requiring the independence of
the tests, providing a lower bound for the significance level.
To achieve a confidence level of $99.76 \%$
with the normal distribution it is necessary to consider a bit more than
3 standard deviations, this is $3.036 \sqrt{\sigma_i^2+\sigma_j^2}$.
In this case, all pairs of exponents seem to be compatible
except one, corresponding to Manus and Darwin sites.
These sites yield a difference between their exponents
equal to $0.055 \pm 0.053$, 
which we can consider in the limit of rejection 
and makes the decision about the universality on the value of exponents
a very critical one. 
In any case, the Bonferroni {and \v{S}id\'ak} corrections seem too generous
(in order to claim for universality,
or too conservative in order to detect differences).
They reduce the type I error (false positive) at the cost
of an enormous increase of the type II error (false negative).
The purpose of these examples {was} to illustrate the difficulty
of dealing with multiple testing.
Multiple tests or global hypothesis testing are important research 
topics in statistics, where significant contributions are currently 
made (see, for instance, Ref. \cite{Marin_Roldan}), 
but there is not a definitive method or solution at the moment.

An additional and more important problem is the identification, on the one hand, of $\sigma_i$, the standard deviation
of the maximum likelihood exponent for fixed $a_i$ and $b_i$ with, on the other hand, the
true uncertainty of the exponent for the whole optimization process.
{It is the latter which should be used in the tests.}
We expect this uncertainty to be larger than the standard deviation estimation,
but its precise value is hard to quantify.
{So, the approaches mentioned above cannot be applied as the uncertainties 
are in fact not known.}
Other, more refined approaches to multiple testing \cite{Benjamini}
face the same problem when the precise quantification
of the uncertainty in individual tests is not possible.
{In any case, if the uncertainties have to be larger it seems clear
that the Bonferroni and \v{S}id\'ak corrections are not able to reject the null
hypothesis that all the values of the exponents are compatible between them.}

\section{Restricted permutational test}
\label{sec_four}

Instead of multiple testing, we propose an alternative track, using a permutational test,
which avoids the drawbacks just mentioned.
{We deal in this section with a very simple test, 
see the next one for a more ellaborated procedure.}
{The null hypothesis $H_0$ is here that for the common range $a\le s \le b$
all exponents are the same, i.e., $\tau_i=\tau_j$, for all $i$ and $j$
(note that before we had $M(M-1)/2$ null hypotheses,
one for each pair).}
What we need first is a statistic that quantifies the divergence
between all the exponents, in such a way that the larger the value
of the statistic, the stronger the evidence against the null hypothesis.

{In order to construct this statistic}
we may take a weigthed sum of $(\hat \tau_i-\hat \tau_j)^2$, 
or of $|\hat \tau_i-\hat \tau_j|$, for all $i$ and $j$,
or rather, the maximum of all the differences, $\max_{\forall i j}(\hat \tau_i-\hat \tau_j)$.
The first option gives more weight to the most extreme differences
than the option of taking instead the {sum of the} absolute values of the differences,
but lest weight than the option of the maximum difference.
So, the second power of the differences 
constitutes a compromise between the importance given
to the extremes and the importance given to the central values.
Our particular selection {for the statistic} is
\begin{equation}
{\widehat{\Theta}}=\displaystyle\sum\limits_{i=1}^{M-1}\sum\limits_{j=i+1}^{M} 
\frac{n_{i}n_{j}}{n_{i}+n
_{j}}
(\hat \tau_{i}-\hat \tau_{j})^2,
\label{teststatistic}
\end{equation}
%
%
{where $n_i$ is the number of data of dataset $i$
in the common power-law range, $a\le s \le b$.
When this statistic refer to the empirical data
(and not to permutations) 
we will call it $\widehat{\Theta}_{\text{data}}$.}

The prefactor depending on $n_i$ and $n_j$ can be easily justified.
Under the null hypothesis, and for independent datasets, 
each $\hat \tau_i-\hat \tau_j$ has zero mean 
and variance 
{equal to the sum of the variances of each exponent.}
But we can assume {that the variance of each exponent} 
is proportional to $1/{n_i}$.
Indeed, when the fitting range is fixed a priori \cite{Aban},
the proportionality constant depends on the value
of the exponent and on the fitting range 
as,
\begin{equation}
\sigma_{i}^2
=\frac 1 {{n_i }}\left[\frac 1 {(\hat \tau_i-1)^2} - \frac{r_i^{\hat\tau_i-1}\ln^2 r_i}
{(1- r^{\hat\tau_i-1})^2}\right]^{-1},
\label{sigma0}
\end{equation}
with $r=a/b$; 
then, for identical exponents
and for a common fitting range, $\sigma_i^2 \propto 1/{n_i}$.
We consider then that when the fitting range is not fixed 
but optimized (as it is our case),
this dependence still holds.
Therefore, under the null hypothesis,
the expected value $\langle (\hat \tau_i-\hat \tau_j)^2  \rangle$
is proportional to $1/n_i + 1/n_j$,
and so,
$ n_i n_j  \langle (\hat \tau_i-\hat \tau_j)^2  \rangle /(n_i+n_j)$
is the same for all $i$ and $j$,
independently of the number of data $n_i$ and $n_j$.
Then,
every term in $\widehat{\Theta}$ has the same expected value
and contributes the same to the sum (on average).
If we did not include the prefactor we would be giving more
weight to the smallest datasets.
On the contrary, if, for some reason, we 
wanted to give more weight to the largest datasets
we could have taken $ [n_i n_j /(n_i+n_j)]^2$ as a prefactor, 
for example.
}

The scale for $\widehat{\Theta}$ is provided by the {\it achieved significance level} or 
$P-$value of the test, which is defined as the probability that, under the null hypothesis $H_0$,  
the random variable $\widehat{\Theta}$ is larger than the value we obtained for 
the observed data $\widehat{\Theta}_{\text{data}}$,
{i.e.,}
 \begin{equation}
P
=\text{Prob}\{\text{}\widehat{\Theta}\geq\widehat{\Theta}_{\text{data}} \text{ $|$ } H_{0}\text{ is true}\},
\elabel{pvalue}
\end{equation}
so, the smaller the $P-$value, the stronger the evidence against $H_{0}$
{(we use capital $P$ in order to distinguish this $P-$value 
from the $p-$value of the power-law fit)}.
{Although each term in the sum of $\widehat{\Theta}$ follows a gamma distribution,
with the same parameters
(as each is the square of a normal variable),
there is no easy way to compute the distribution of the sum,
due to the fact that the terms are not
independent.
That is, even if all datasets $i,j,k$, etc.,
are independent, the terms $(\hat \tau_i-\hat \tau_j)^2$, $(\hat \tau_j-\hat \tau_k)^2$, etc., are not. 
}

Fisher's permutation test \cite{EfronTibshirani1993} (also called randomization test) 
is a clever way to compute the $P-$value in cases like these. 
It is based on the idea that, if the null hypothesis is correct,  
any data value could correspond to any dataset, and the data values 
(the size of the rain events in our case) are therefore interchangeable. 
{In order to proceed with the test, we combine}
the $n_{1}+n_{2}+...+n_{M}$ 
observations in the common power-law range
into a single meta-dataset and take $M$ random samples of sizes 
$n_{1}$, $n_{2}$, ..., $n_{M}$ without replacement
(this is done just by a permutation or reshuffling of the meta-dataset,
and then taking consecutive $n_i$ values). 
This generates $M$ new datasets with the same number of data than the initial ones.

Next, we fit the power-law exponents {(in the common fitting range)} 
for each of the $M$ permuted or reshuffled datasets
and from their values we compute the new test statistic $\widehat{\Theta}_{\text{sh}}$,
in the same way as for $\widehat{\Theta}_{\text{data}}$
(sh stands for shuffled now).
As the fitting range, given by $a$ and $b$ is fixed,
the fit of the exponent is simple, using just maximum likelihood estimation
(neither goodness-of-fit Kolmogorov-Smirnov tests nor
simulations are necessary).
The distribution of the test statistic, under the null hypothesis, 
is obtained repeating the permutation process a large enough number of times,
$N_{\text{sh}}$.
{In our case, we always take $N_{\text{sh}}=100$.}
 With that we can compute easily an approximation of the $P-$value by
  \begin{equation}
P\approx
\frac
{\#{\{\widehat{\Theta}_{\text{sh}} \geq \widehat{\Theta}_{\text{data}}\}}}
{N_{\text{sh}}}
\label{pvalueaprox}
\end{equation} 
where $\#{\{\widehat{\Theta}_{\text{sh}} \geq \widehat{\Theta}_{\text{data}}\}}$
is the number of permutations for which 
$\widehat{\Theta}_{\text{sh}} \geq \widehat{\Theta}_{\text{data}}$.
Notice that this approach does not make use 
of the errors of the exponents (which is an advantage).
For our $M=7$ rain datasets 
we obtain $\widehat{\Theta}_{\text{data}}=30.3$, 
which, after the permutational procedure, leads to 
$P=0.03$.
So, at the $5 \%$ significance level, we reject the hypothesis
that all the exponents take the same value and
we cannot give statistical support to universality in rainfall.



\section{Complete permutational test}
\label{sec_five}

One can realize that the previous procedure has at least one drawback.
The choice of a fix common fitting range seems somewhat artificial,
due to the fact that this range is optimum for some empirical dataset 
but not necessarily for 
any of the permutations, a fact that can introduce a bias in the procedure.
In other words, the fit can be better for the true datasets than 
for the reshuffled ones, and in this way we are not treating the latter
in the same way as the former. This is something that needs to be avoided;
simulated or permuted data have to be treated in exactly the same way
as the real data to avoid biases and artifacts \cite{ClausetShaliziNewman2009,Malgrem}.


In order to proceed in the same way with the permuted data, we have to look, for each
reshuffled dataset, for the most appropriate fitting range, and then select the 
common power-law range.
Thus, 
{we introduce a modification of the test in which}
we do not reshuffle the common part of the data in which all distributions 
are power law, but we reshuffle the whole data.
That is, we aggregate the $N_1+N_2 + \dots + N_M$ data,
where $N_i$ is the total size of dataset $i$, and take random samples, 
without replacement, of size $N_1$, $N_2$, $\dots$, $N_M$,
and, we insist, we perform with these datasets in the same way as with the true data. 
Notice that in the previous subsection the null hypothesis was
that, over a common range given by $a$ and $b$, all the distributions were power laws with the same exponent.
Now the null hypothesis is different, rather, we test if there exist a common range over which 
all the distributions are power laws with the same exponent, 
but we do not specify which is that common range.

The procedure is summarized as follows, {for every permutation $\ell$
(with $\ell$ from $1$ to $N_{\text{sh}}$)}:
\begin{enumerate}
\item For each reshuffled data set, $i=1,\dots M$, calculate the 
values of $a_i^{(\ell)}$ and $b_i^{(\ell)}$ which lead to the largest number of data 
in a power law fitted in that range, provided that $p \ge 0.10$.
As already mentioned, 
this is done by maximum likelihood estimation of the exponent 
plus the Kolmogorov-Smirnov test plus Monte Carlo simulations
of a power law in the range $a_i^{(\ell)} \le s \le  b_i^{(\ell)}$.

\item Select the common fitting range for the $M$ reshuffled datasets, 
as $a^{(\ell)}= \max_{\forall i} a_i^{(\ell)}$ and $b^{(\ell)}= \min_{\forall i} b_i^{(\ell)}$
({and verify that $a^{(\ell)} \ll b^{(\ell)}$}). 

\item Calculate new exponents $\hat \tau_i^{\ell}$ in the common fitting range, by maximum likelihood 
estimation alone. 

\item Calculate the test statistic $\widetilde{\Theta}_{\text{sh}}^{(\ell)}$ 
(with a new definition, see Eq. (\ref{teststatistic2}) below).

\end{enumerate}
{And the same is repeated for every permutation.
Then, the $P-$value is calculated as in the previous case, Eq. (\ref{pvalueaprox}).}
All steps, from 1 to 4, are exactly the same as for empirical data.


Note that  
step 1, the most
time consuming (due to the Monte Carlo simulations), 
was removed in the method of the previous section, 
as the common fitting range was the same in all permutations
(obviously, step 2 was also unnecessary). 
This meant that only data inside the common fitting
range had to be permuted.
Now, we release such a restriction, and
as a result, each collection of reshuffled datasets will lead to a
different common fitting range.
In this case, one has to take care in order to compare the test statistic
$\widetilde{\Theta}$ corresponding to the reshuffled data and the real data, 
as our previous definition, Eq. (\ref{teststatistic}), 
did not take into account that different fitting ranges may
correspond to different variances of $\hat \tau_i-\hat \tau_j$.
Indeed, 
from Eq. (\ref{sigma0}) 
we know that
{the standard deviation of the estimation of the power-law exponent $\hat \tau_i$
(for a fixed fitting range)
will depend not only on the number of data in the power-law range $n_i$
but also on the ratio of the cut-offs, $r=a/b$.}

In general, Eq. (\ref{sigma0}) teaches us that the larger the fitting range, the smaller $r$,
and the smaller also $\sigma_i$ (even if the number of data keeps constant).
In order to compensate this fact, we define the test statistic directly
as
\begin{equation}
\widetilde{\Theta}_{\text{data}}
=\displaystyle\sum\limits_{i=1}^{M-1}\sum\limits_{j=i+1}^{M} 
\frac{
(\hat \tau_{i}-\hat \tau_{j})^2
}{( \sigma_i^2 + \sigma_j^2)},
\label{teststatistic2}
\end{equation}
using the notation for the empirical data,
with $\sigma_i$ referring to the standard deviation
in the fixed fitting range case, see Eq. (\ref{sigma0}).
For the reshuffled data, an analogous definition 
yields $\widetilde{\Theta}_{\text{sh}}$,
but note that the resulting exponents $\hat\tau_i^{(\ell)}$
and standard deviations $\sigma_i^{(\ell)}$ will be different to the ones 
of the first test, as the ranges are different (see previous section).
By dividing by the sum of the variances we do not
only ensure that each pair of datasets contributes the same
to the statistic (on average)
but also
that the statistic has the same average value
for each permutation (under the null hypothesis). 
If the fitting range were the same for all permutations,
this statistic would be essentially the same
(except for a constant factor)
as the one employed in the previous section, 
Eq. (\ref{teststatistic}).
Note nevertheless that, in contrast with the multiple testing
explained before, the outcome of this test is not influenced
by the size of the ``error bars'' associated to the exponents, 
i.e., we could duplicate the value of all the $\sigma_i$
and the $P-$value of the test would not change.
In other words, we just use $\sqrt{\sigma_i^2+\sigma_j^2}$ as
a scaling factor of the differences between the exponents.
The results of this generalized test for the $M=7$ data yield 
$\widetilde{\Theta}_{\text{data}}=44.8 $ and 
$P=0.04$;
again, the null hypothesis of universality can be rejected
at the $5 \%$ significance level.

\section{Permutational test without a common power-law range}
\label{sec_six}

A criticism to our previous permutational methods is that the restriction of the fits 
to a common range reduces considerably the {number of data}, 
which increases
the fluctuations of the exponents, making more difficult to detect
differences between them. 
So, although the methods have shown powerful enough in our particular
case, they could fail to detect true differences in other problems.
%
In fact, as in the test of the previous section we have corrected
for the different fitting ranges of each dataset and each permutation, 
one can realize that we would not need 
to look for a common fitting range in any case and we could suppress
steps 2 and 3 in the procedure
(both for the permuted data and for the real data).
This procedure has the advantage that we use the complete power-law ranges
of each dataset.
Nevertheless, the permutation of the whole datasets leads to 
shorter power-law ranges, as one only would expect
strict power-law behavior over the common power-law range.
So, this variation of the test is limited also by the shortness of the power-law ranges
in the reshuffled datasets. 


%

{A better option is to transform the data in order that all of them
are defined in the same range, but without disregarding any part
of the power-law portion of the data.
Remember that in the first version of the test we 
reshuffled only data in the common power-law range, $a \le s \le b$,
whereas in the second one we reshuffled the whole data, $0 \le s \le \infty$.
We pretend now to reshuffle the data keeping for each data set $i$
its own power-law range, $a_i \le s \le b_i$.
In order to do so, we transform each dataset in such a way that if 
it is power-law distributed between $a_i$ and $b_i$ 
with exponent $\bar \tau_i$,
its distribution turns out to be
uniform between 0 and 1.
This is simply done by replacing each size $s$ by a value $u$ given by
\begin{equation}
u=S_i(s)= \frac{s^{1-\bar \tau_i}-b_i^{1-\bar \tau_i}}{a_i^{1-\bar \tau_i}-b_i^{1-\bar \tau_i}},
\label{uSis}
\end{equation}
where remember that $\bar \tau_i$ is the 
exponent of dataset $i$ in its complete power-law range, 
$a_i \le s \le b_i$ and $S_i(s)= \int_s^{b_i} f_i(s')ds'$
is the complementary cumulative distribution function
(or survivor function) of a truncated power-law distribution
defined between $a_i$ and $b_i$.
So, each dataset is transformed in this way, with its own
exponent $\bar \tau_i$ and range, and then the resulting values
are reshuffled. 
After this, each reshuffled set is transformed back
to the original form, 
\begin{equation}
s=S_i^{-1}(u)=\frac 1 {\sqrt[\bar \tau_i-1]
{b_i^{1-\bar \tau_i}+(a_i^{1-\bar \tau_i}-b_i^{1-\bar \tau_i})u}},
\end{equation}
where this equation is just the inversion of Eq. (\ref{uSis}).
If $u$ is uniformly distributed between 0 and 1, 
this yields a power law defined between
$a_i$ and $b_i$ with exponent $\bar\tau_i$,
but note that the concrete data values are different from the original ones,
due to the reshuffling. 
{This is a procedure to obtain resampled data with the same
exponents $\bar \tau_1, \bar \tau_2, \dots \bar \tau_M $}
than the original ones, provided that all original datasets
were indeed power-law distributed with those exponents.

Now, for each permuted data set, in order to perform the fit and the goodness-of-fit test
 we only need to apply 
step 1 of the usual procedure
(no common range is necessary anymore).
Note that the resulting cut-offs should verify $a_i^{(\ell)} \ge a_i$
and $b_i^{(\ell)} \le b_i$
(as our current data is only defined between $a_i$ and $b_i$).
The test statistic for the empirical data is calculated 
from Eq. (\ref{teststatistic2}),
but replacing $\hat \tau_i$ by $\bar\tau_i$
and using the corresponding value of the standard deviations.
For the reshuffled data the test statistic is analogous. 
The outcome of the method is 
$\widetilde{\Theta}_{\text{data}}=82.6 $ 
with $P=0.01$ and therefore the null hypothesis
of universality in the value of the exponents
is clearly rejected.
}



\section{Discussion and conclusions}

{We may ask about the reasons behind the rejection of the universality hypothesis. Looking at Table \ref{table_comparison}, it is clear that the largest differences between pairs of exponents, in terms of their standard
deviations, always involve the Darwin site. 
Indeed, the value of the exponent for this site does not seem compatible with the values for Manus, Heselbach, and Graciosa. 
But remember that an analysis derived from Table \ref{table_comparison} 
is not reliable, due to the problems of multiple testing and to the fact that
the standard deviation of the exponents are not well determined.

We may use the permutational tests to detect if the rejection of universality
is associated to the Darwin site, just disregarding this site and repeating
the permutational test for the remaining six sites. 
The outcome is   
a $P-$value larger than 0.05 for the  
simple restricted permutational test,
and larger than 0.10 for the other two more advanced tests,
signaling in any case that universality cannot be rejected when the Darwin site is excluded.

In order to be more general, and to avoid biases, we may try to detect if the rejection of universality can be associated to other sites, 
so we systematically repeat the permutational tests for all the combinations
of six sites (i.e., removing another site instead of Darwin). 
The results displayed in Fig. 1 show that in these alternative cases, 
were the Darwin site is kept and other sites are removed, 
universality is rejected in most cases.
The main exception is the Manus site, 
which yields $P-$values larger than 0.05 (but smaller than 0.10)
for the second and third testing procedure.

\begin{figure}
{\includegraphics[width=0.90\textwidth]{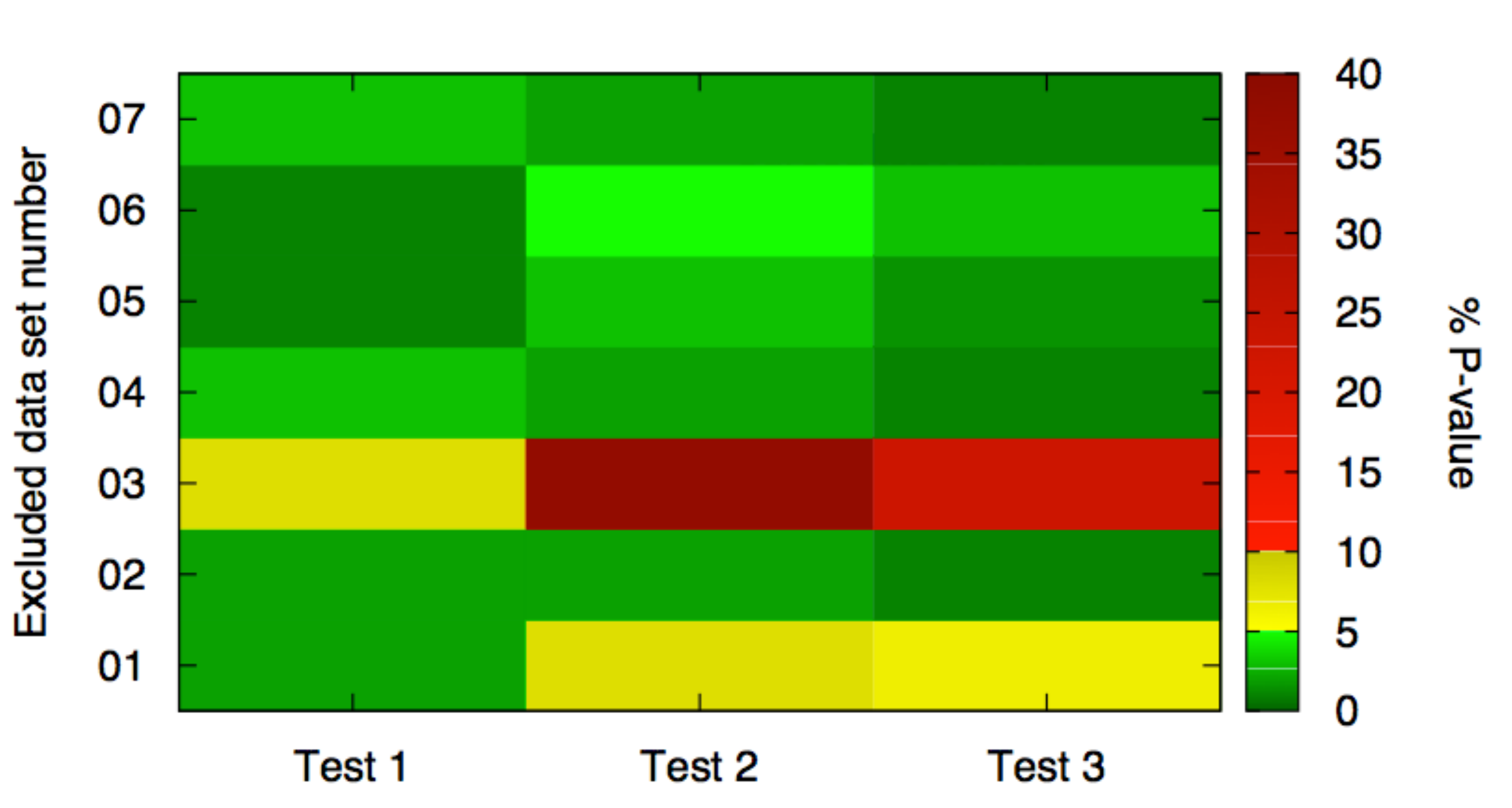}} 
\caption{\label{fig1}
$P-$values for the three permutational tests when one of the sites,
from 1 to 7, is eliminated. Sites are labeled as in Table \ref{table_exponents}, 
with Darwin corresponding to site 3. Low $P-$values mean that violation of universality
for the values of the critical exponents is significant. 
Although $3\times 7$ tests are performed, no correction for multiple testing is
taken into account.
}
\end{figure}
 
So, we conclude that the reason of the rejection
can be attributed to a unique site, which is Darwin, but this attribution
is purely statistical, as we cannot find any physical or technical 
reason why this site could be problematic (in contrast to the sites excluded in Ref. \cite{PetersDelucaETAL2010}.).
Alternatively, the same argument could be applied to Manus,
although with less strength in this case, 
because of the moderately large $P-$values obtained when
this site is removed instead of Darwin.
In consequence, the only clear statement is that 
universality is rejected essentially because of the incompatibility
of the exponents for Darwin and Manus.

Another factor to take into account is the uncertainty in the measurement of the rain rate. 
It could be that, in principle, universality holds, but experimental errors make
the exponents slightly but significantly different, in such a way that universality
appears as violated.
We have no access to the value of the true rain rates (without experimental errors), but considering that the accuracy of the pluviometers is established to be 5 \%
for the minutal rain rate measurements 
\cite{org_brochure},
we consider the effect of this error in the measured values, 
adding to the measured values of the rate a random noise with a standard
deviation equal to the 5 \% of the measured rate.
The analysis of these noisy data shows no significant difference 
with the original results, rather, the difference is smaller
than the associated errors, so we conclude that this effect seems to be too
small to be the cause of the lost of universality.  We do not perform further tests 
as we consider highly unlikely that systematic biases are affecting our results. 
}

{In summary},
we have developed permutational tests to deal with the universality 
or not of the critical exponents arising from power-law probability
distributions. More common methods require the precise estimation 
of the uncertainty of the exponents, which is difficult for
our sophisticated fitting and testing procedure of the value
of the exponents \cite{Corral_Deluca}.
Moreover, those common methods suffer from the difficulties of multiple testing,
as for instance the artificially high values of the familywise 
significance level and the non-independence of the tests.

In the case of rain-event size distributions
our alternative permutational tests give clear and unambiguous results:
despite the fact that the differences between the exponents are rather small,
the universality hypothesis is rejected, both for a extremely simple version 
of the permutational test and for two more complete implementations that avoid
some artificial biases in the procedure.
The last of these is able to use the complete power-law range of the datasets. 
{Breaking of universality can be attributed to the Darwin site.}

{If asked about which of the tests is better,
the first one, the restricted permutational test (Sec. \ref{sec_four}),
has the clear advantage of a minimal computational load, 
but we have already discussed that can be biased.
The second test (Sec. \ref{sec_five}) avoids some bias of the first one, 
so it is better if the number of data does not make its computational cost prohibitive.
Nevertheless, both tests are restricted to a common fitting range,
which should not be a problem 
if the fitting ranges of the different datasets are not
very different from each other; however, in the opposite case, 
the number of data that is disregarded is too high, 
and it is even possible that a common fitting range does not exist,
in such a case we recommend the third testing method (Sec. \ref{sec_six}).
In general, the third method has a lower computational
cost than the second one, so, we always recommend the use
of the third method with respect the second one. 

We have also verified that this third method yields
a nearly uniform distribution of $P-$values when used in a control
case when all exponents are the same, by construction.
Also, the method is powerful enough to detect a change of 0.10 in one of the exponents, keeping the rest identical, in particular when the change in the exponent corresponds to the largest dataset.

{
In this paper
we have seen how
relatively involved permutational tests 
are necessary in order to test the hypothesis
of universality in critical exponents of probability distributions.
The complications 
arise from the fact that we fit truncated power laws to the data 
(i.e., power laws with an upper cut-off $b_i$). 
Similar difficulties would arise for power-law distributions
with exponential-like tails, as the ones in Ref. 
\cite{Serra_Corral}.
The problems 
are substantially reduced when the power laws are not truncated from above
(i.e., $b_i=\infty$); 
in this case, it is enough to rescale
every value of the random variable $s$ by the lower cutoff $a_i$
of the corresponding dataset $i$, so that
the random variable of interest becomes $s/a_i$, which
are defined in the common range $[1, \infty)$.
Then, the simple restricted permutational test
of Sec. \ref{sec_four} should suffice to test universality.
}
}

It is worth mentioning also that for the original data analyzed in 
Ref. \cite{PetersDelucaETAL2010} (which contain less events due to 
the shorter time span covered there), the first and second tests are not able to find 
deviations from universality, but the last test is. 
So, it is when a critical amount
of data is available that these deviations show up more clearly.
This issue needs to be further investigated, but one could speculate that
the deviations
could arise from a lack of stationarity, due to long-term slight variations
associated for instance with El Ni\~no phenomenon or to seasonal fluctuations.
At present, the number of data available is still small to shed light on this point.
On the other hand, even if the existence of a universal mechanism for atmospheric convection would be a useful approximation to model this very complex phenomena, 
changes under climate change forcing could also introduce non-trivial adjustments in the associated parameters. This would be again 
extremely  challenging to detect with the current data availability. 

\section*{Acknowledgements}
The authors started their research on rain thanks to O. Peters.
Data were obtained from the Atmospheric Radiation Measurement Program
sponsored by the US Department of Energy, Office of Science, Office of Biological and
Environmental Research, Environmental Sciences Division.
Research expenses were founded by grants 
FIS2009-09508, from the disappeared Spanish MICINN, 
FIS2012-31324 and MTM2012-31118, from Spanish MINECO, 
2009SGR-164 and
{2014SGR-1307,}
from AGAUR,
and 
UNAB10-4E-378 from ERDF ``A way of 
making Europe''.

%

\begin{thebibliography}{10}
\bibitem{footnote}
For instance, at a significance level of 5 \% the uncertainty of $\hat \tau_i$
  is given by $\pm z_c \sigma_i$, with $z_c=1.96$. {If}
  $\sigma_1=\sigma_2=\sigma$, the probability of $|\hat \tau_1 - \hat \tau_2| >
  \sqrt{2} \sigma z_c$ is indeed $\text{Prob}\{|z|>z_c\}=0.05$, but testing if
  $|\hat \tau_1 - \hat \tau_2| > 2 \sigma z_c$ reduces drastically the
  significance level to $\text{Prob}\{|z|>\sqrt{2}z_c\}= 0.0066$.
\end{thebibliography}
%

\end{document}